\documentclass[prx,twocolumn,amsmath,amssymb,superscriptaddress,longbibliography]{revtex4-2}

\usepackage[pdftex]{graphics}
\usepackage{graphicx}
\usepackage{times}

\graphicspath{{Figures/}}

\usepackage{xcolor}
\usepackage{color}
\usepackage[colorlinks=true, urlcolor=blue, linkcolor=blue, citecolor=blue, pdftex]{hyperref}

\definecolor{darkblue}{HTML}{004D6B}
\definecolor{darkred}{HTML}{8c1515}
\definecolor{darkgreen}{HTML}{006400}
\usepackage{hyperref}
\hypersetup{
	colorlinks=true,
	urlcolor=darkred,
	citecolor=darkblue,
	linkcolor=darkred,
	breaklinks
}

\begin{document}

\title{RIXS observation of bond-directional nearest-neighbor excitations 
	in the Kitaev material Na$_2$IrO$_3$}

\author{M. Magnaterra}
\author{K. Hopfer}
\affiliation{Institute of Physics II, University of Cologne, 50937 Cologne, Germany}
\author{Ch. J. Sahle}
\affiliation{ESRF, The European Synchrotron, 71 Avenue des Martyrs, CS40220, 38043 Grenoble Cedex 9, France} 
\author{M. Moretti Sala}
\affiliation{Dipartimento di Fisica, Politecnico di Milano, I-20133 Milano, Italy}
\author{G. Monaco}
\affiliation{Dipartimento di Fisica e Astronomia "Galileo Galilei", Universit\`{a} di Padova, Padova, Italy}
\author{J.~Attig}
\author{C. Hickey}
\affiliation{Institute for Theoretical Physics, University of Cologne, 50937 Cologne, Germany}
\author{I.-M. Pietsch}
\author{F. Breitner}
\author{P.~Gegenwart}
\affiliation{Experimental Physics VI, Center for Electronic Correlations and Magnetism, University of Augsburg, 86159 Augsburg, Germany}
\author{M. H. Upton}
\author{Jungho~Kim}
\affiliation{Advanced Photon Source, Argonne National Laboratory, Argonne, Illinois 60439, USA}
\author{S. Trebst}
\affiliation{Institute for Theoretical Physics, University of Cologne, 50937 Cologne, Germany}
\author{P. H. M. van Loosdrecht}
\affiliation{Institute of Physics II, University of Cologne, 50937 Cologne, Germany}
\author{J. van den Brink}
\affiliation{Institute for Theoretical Solid State Physics, IFW Dresden, 01069 Dresden, Germany}
\affiliation{Institute for Theoretical Physics and W\"urzburg-Dresden Cluster of Excellence ct.qmat, Technische Universit\"at Dresden, 01069 Dresden, Germany}
\author{M. Gr\"{u}ninger}
\affiliation{Institute of Physics II, University of Cologne, 50937 Cologne, Germany}

\date{January 19, 2023}

\begin{abstract}
Spin-orbit coupling locks spin direction and spatial orientation and generates, in 
semi-classical magnets, a local spin easy-axis and associated ordering.
Quantum spin-1/2's defy this fate: rather than spins becoming locally anisotropic, 
the spin-spin interactions do.
Consequently interactions become dependent on the spatial orientation of bonds 
between spins, prime theoretical examples of which are Kitaev magnets. 
Bond-directional interactions imply the existence of bond-directional magnetic modes, 
predicted spin excitations that render crystallographically equivalent bonds 
magnetically inequivalent, which yet have remained elusive experimentally.
Here we show that resonant inelastic x-ray scattering allows us to explicitly probe 
the bond-directional character of magnetic excitations. 
To do so, we use a scattering plane spanned by one bond and the corresponding 
spin component and scan a range of momentum transfer that encompasses multiple 
Brillouin zones. Applying this approach to Na$_2$IrO$_3$ we establish the 
different bond-directional characters of magnetic excitations at $\sim$\,10\,meV 
and  $\sim$\,45\,meV.\@ Combined with the observation of spin-spin correlations 
that are confined to a single bond, this experimentally validates the Kitaev character 
of exchange interactions long proposed for this material. 
\end{abstract}

\maketitle

%%%%%%%%%%%%%%%%%%%%%%%%%%%%%%%%%%%%%%%%%%%%%%%%%%%%%%
% Introduction
%%%%%%%%%%%%%%%%%%%%%%%%%%%%%%%%%%%%%%%%%%%%%%%%%%%%%%

The physics of quantum magnets with bond-directional interactions can be captured by so-called 
compass models \cite{Nussinov15}, quantum theories of matter in which the couplings between 
different spin components are inherently spatially (typically, direction) dependent.
Theoretically, this class of models harbors a range of interesting emergent physical phenomena, 
including the frustration of (semi-)classical ordered states on unfrustrated lattices, and 
a boost of quantum effects, prompting, in certain cases, the appearance of 
quantum spin liquids \cite{Broholm20,Savary16,Knolle19} -- Kitaev spin liquids are 
well-known examples \cite{Kitaev06,Hermanns18}.
In spin-1/2 materials, spin-orbit coupling naturally induces bond-directional 
spin-spin interactions. These can dominate when spin-orbit coupling becomes large, 
e.g., in 4$d$ and 5$d$ transition metal compounds \cite{Jackeli09,Rau2016,Trebst22}.
However it has remained a principal challenge to experimentally identify the 
fingerprints of bond-directional magnetic interactions \cite{Chun15} and 
to establish methods to systematically explore their consequences for 
elementary magnetic properties.

Here we show that bond-directional {\it excitations} (BDE) -- spin excitations that 
render crystallographically equivalent directions magnetically inequivalent
-- are a hallmark of bond-directional magnetic {\it interactions} and demonstrate 
how to use resonant inelastic x-ray scattering (RIXS) to directly probe these BDE.\@
The challenge to resolving the bond-directional character of magnetic modes is that 
it requires simultaneous knowledge of both the spin operator creating the excitation 
and the direction of the bond involved. 
We introduce a RIXS geometry that yields the former via the polarization dependence 
and the latter by measuring across {\em multiple} Brillouin zones. Using this method 
on Na$_2$IrO$_3$, a Kitaev material exhibiting a proximate spin liquid regime 
\cite{Banerjee16,Mehlawat17,Revelli20}, we establish the different bond-directional 
characters of spin-conserving and spin-flip excitations at $\sim$\,10\,meV 
and $\sim$\,45\,meV, respectively.

\begin{figure}[t]
\centering
\includegraphics[width=\columnwidth]{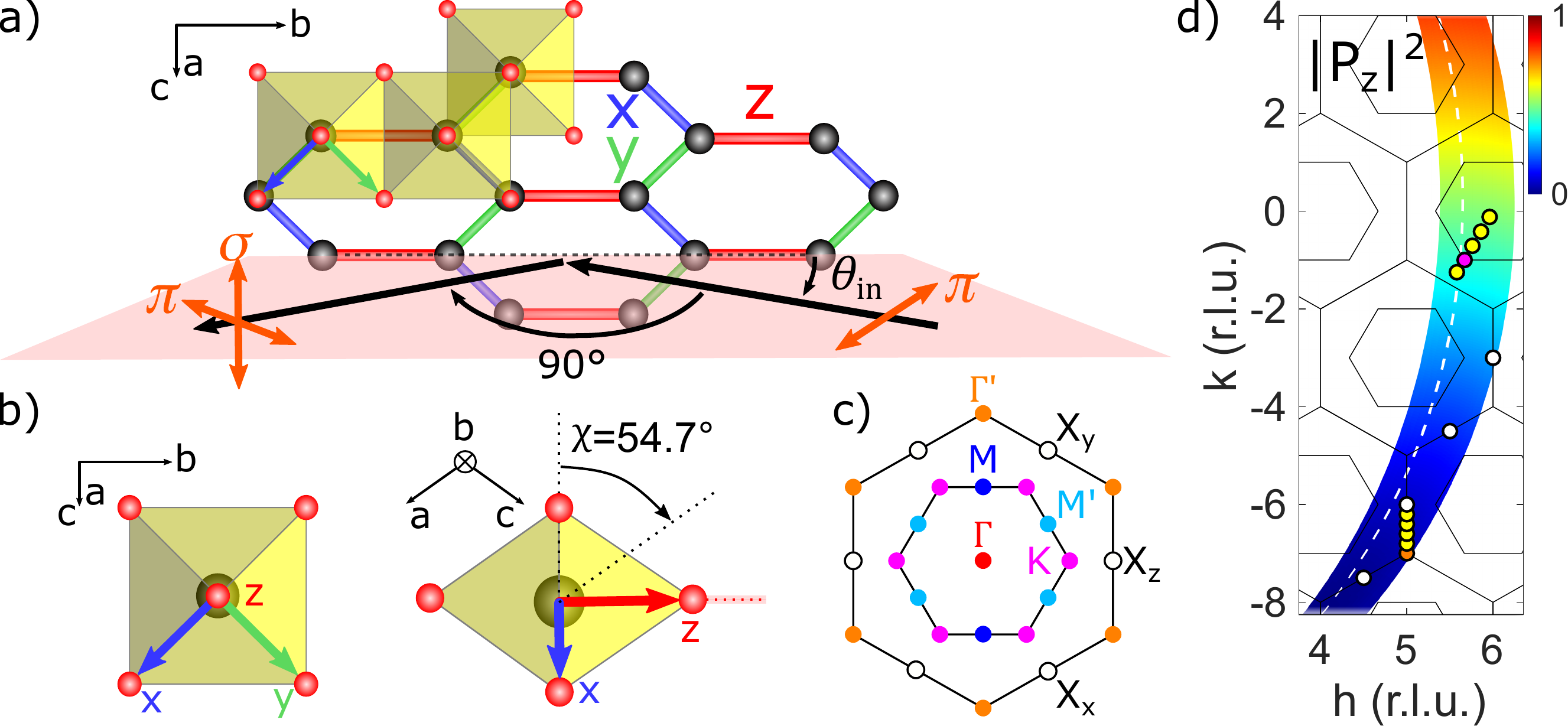}
\caption{\textbf{(a) Sketch of scattering geometry with tilted sample.} 
	We use incident $\pi$ polarization while the RIXS intensity is summed over 
	outgoing $\pi$ and $\sigma$ polarizations. 
	(b) Front and side views of a single IrO$_6$ octahedron. For a tilt angle 
	$\chi$\,=$54.7^\circ$, $j^z$ lies in the scattering plane.
    (c) 2D high-symmetry points in the first and second Brillouin zones around $\Gamma$. 
    A $\Gamma$ point occurs for $h+k$ even and $k$\,=\,$3m$ with integer $m$ and $h$, 
    see (d). BDE are most evident around $X_\gamma$. 
	(d) Polarization factor $|P^z(\mathbf{q})|^2$ across the covered $\mathbf{q}$ range.
	The dashed line corresponds to scanning $\theta_{\rm in}$ from $5^\circ$ to 
	$65^\circ$ for $\chi$\,=\,$54.7^\circ$ and scattering angle $2\theta$\,=\,$90^\circ$.  
    Experimentally, we vary $\chi$ between $52^\circ$ and $64^\circ$. 
	Symbols mark the $\mathbf{q}$ points addressed in Fig.~\ref{fig:spectra}. 
	For the high-symmetry points, symbol colors refer to the sketch in c).
} 
	\label{fig:setup}
\end{figure}

Resolving BDE with RIXS on Na$_2$IrO$_3$ is based on polarization selection rules 
but a polarization analysis of the scattered x-rays is not available at the commonly 
used Ir $L$ edge. We have solved this problem by exploiting a {\em tilted} sample 
geometry which highlights polarization effects over the large range of $\mathbf{q}$ 
space that can be covered with hard x-rays, see Fig.~\ref{fig:setup}. 
Tilting the sample puts the spin-orbit entangled $j^z$ component of the  local pseudo-spin 
$j$\,=\,$1/2$ moments into the scattering plane. 
This allows us to disentangle excitations of $j^z$ from those of $j^x$ or $j^y$. 
The anisotropy in $j$ space translates into a characteristic $\mathbf{q}$ dependence 
of the intensity that provides a direct signature of bond-directional behavior.

%%%%%%%%%%%%%%%%%%%%%%%%%%%%%%%%%%%%%%%%%%%%%%%%%%%%%%
% Bond-directional interactions and excitations
%%%%%%%%%%%%%%%%%%%%%%%%%%%%%%%%%%%%%%%%%%%%%%%%%%%%%%

{\it Bond-directional character:}
To illustrate the conceptual relation between bond-directional interactions and BDE 
we consider a central spin on site $i$ surrounded by sites $j$, connected by bonds 
$\gamma_{\langle ij \rangle}$ along crystallographically equivalent directions. 
When exchange interactions are bond-directional, the magnetic Hamiltonian $H_{\gamma}$ 
for two spins on bond $\gamma$ has the property that in general 
$H_{\gamma}$\,$\neq$\,$H_{\gamma \prime}$ even if bonds $\gamma$ and $\gamma \prime$ 
are equivalent from a structural point of view.  
When one now creates a magnetic excitation by perturbing the central spin by operator 
$\hat{O}_i$, the commutator $\left[H_{\gamma {\langle ij \rangle}}, O_i\right]$ 
in general depends on $\gamma$. As a consequence this magnetic excitation distributes 
{\em unevenly} over the crystallographically equivalent bonds, breaking lattice symmetry.\\

An elementary example in which BDE emerge is the honey\-comb Kitaev model with 
$H_{\gamma}$\,=\,$K\,S_i^\gamma S_j^\gamma$ where $\gamma$\,=\,$x,y,z$ simultaneously denotes 
the three nearest-neighbor bonds {\it and} the three spin components, signifying the 
bond-directional character of the interaction. A local operator $S_i^x$ commutes with 
$H_{x}$ but not with $H_{y}$ and $H_{z}$ on the other two bonds, so that perturbing 
the system by $S_i^x$ renders the three crystallographic bonds inequivalent and produces BDE.\@
In particular $S_i^x$ creates localized flux excitations on the two hexagons that share 
the $x$ bond, while the third hexagon connected to site $i$ remains unaffected, 
manifestly breaking the three-fold rotational symmetry of the flux distribution 
and consequently the Majorana modes scattered by the fluxes.

%%%%%%%%%%%%%%%%%%%%%%%%%%%%%%%%%%%%%%%%%%%%%%%%%%%%%%
% RIXS setup and measurements
%%%%%%%%%%%%%%%%%%%%%%%%%%%%%%%%%%%%%%%%%%%%%%%%%%%%%%

Single crystals of Na$_2$IrO$_3$ were grown following the procedure described in \cite{Singh10} 
with 10\,\% extra Ir at 1323\,K for 14\,days. To establish the presence of BDE 
we measured RIXS at the Ir $L_3$ edge at beamline ID20 at the ESRF \cite{Moretti13,Moretti18}. 
The incident energy $E_{\rm in}$\,=\,11.2145\,keV resonantly enhances magnetic 
excitations of the $j$\,=\,$1/2$ moments. The sample surface is parallel to the 
honeycomb plane, i.e., the $ab$ plane. The $b$ axis contains the $z$ bond and 
lies in the horizontal scattering plane such that the incident beam is parallel 
to $b$ for a vanishing angle of incidence $\theta_{\rm in}$, see Fig.~\ref{fig:setup}. 
With a tilt of the sample of $\chi$\,=\,54.7$^\circ$ around $b$, 
where $\chi$ is the angle between the vertical and the honeycomb $ab$ plane,
$j^z$ is lying in the scattering plane. 
The elastic contribution due to Thomson scattering is almost fully suppressed 
by strictly using a scattering angle $2\theta$\,=\,90$^\circ$, 
where all outgoing polarization contributions are perpendicular to the incident 
$\pi$ polarization.
The resolution $\delta \mathbf{q}$ of the transferred momentum equals about 
$(\pm 0.05\,\,\pm 0.1\,\,\pm 0.05)$ reciprocal lattice units (r.l.u.) 
using a 60\,mm iris on the $R$\,=\,2\,m Si(844) spherical diced analyzer. 
The energy resolution is $\delta E$\,=\,25\,meV.\@ 
All RIXS data were corrected for self-absorption \cite{Minola15}.

\begin{figure}[t]
\centering
\includegraphics[width=\columnwidth]{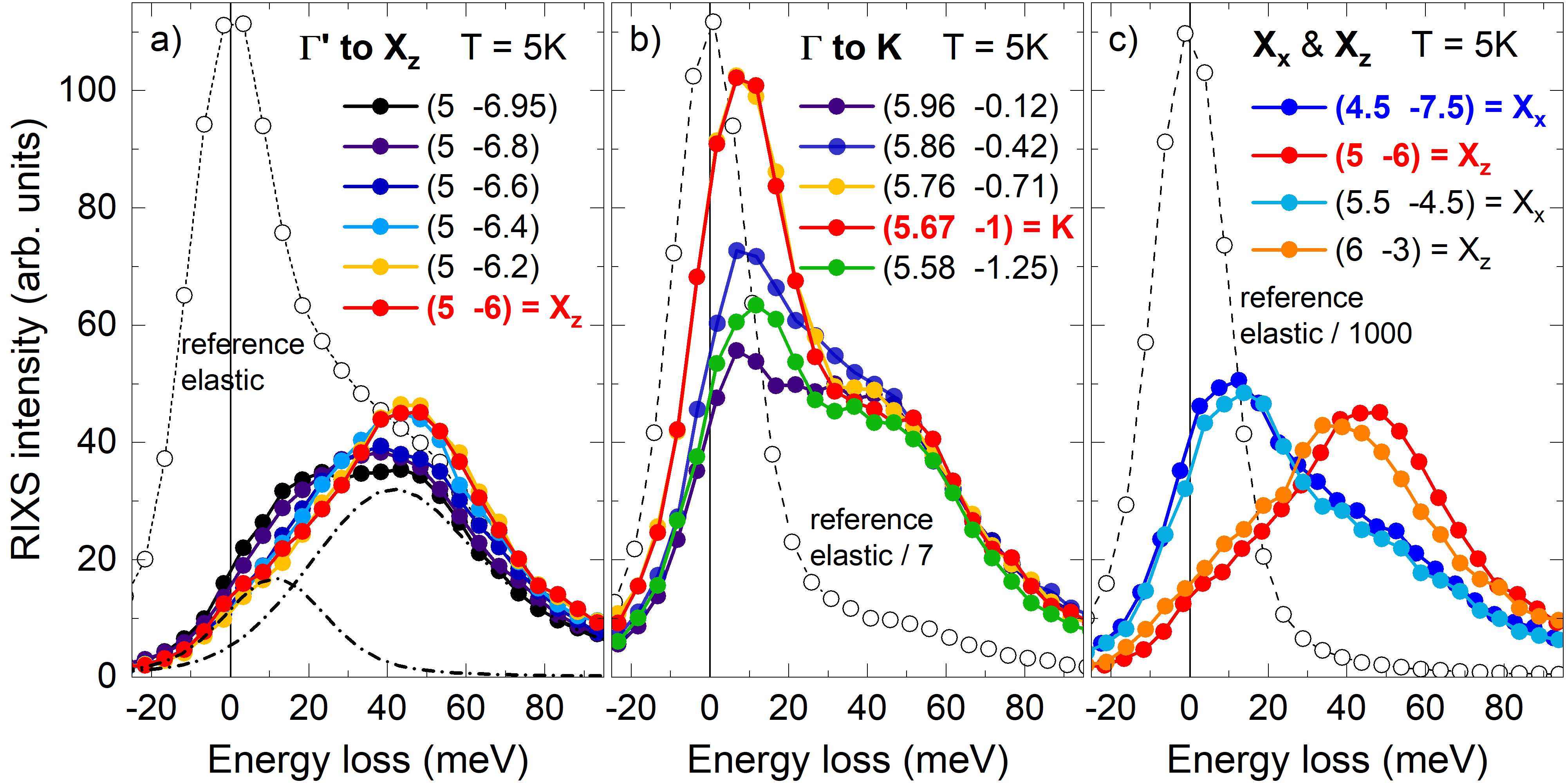}
\caption{\textbf{RIXS spectra at 5\,K at selected q points.}  
    (a) $\Gamma^\prime$ to $X_z$, (b) $\Gamma$ to $K$, (c) $X_x$ and $X_z$, 
    cf.\ Fig.~\ref{fig:setup}. 
	The spectra show two peaks at about 10\,meV and 45\,meV which we attribute to predominantly local nearest-neighbor excitations. 
	Dash-dotted lines in (a) distinguish the two peaks at (5\,\,-6.95).
	Reference spectra recorded with $2\theta$\,$\neq$\,$90^\circ$ and a corresponding large elastic line mark zero energy loss (open symbols, scaled down as indicated). 
	In (c), the different intensities of the two peaks at $X_x$ and $X_z$ provide 
	direct evidence of BDE.
}
\label{fig:spectra}
\end{figure}

RIXS spectra measured at 5\,K at selected $\mathbf{q}$ points show two inelastic 
features peaking around 10\,meV and 45\,meV, see Fig.~\ref{fig:spectra}. 
Their magnetic character is demonstrated by the resonance behavior measured at 
30\,K, i.e., above the 3D ordering temperature $T_N$\,=\,15\,K, 
see \textit{Supplementary Information} \cite{SI}. 
Our focus is on the bond-directional character of excitations expected for a 
Kitaev material, not on low-energy magnons of the ordered state. 
In previous RIXS studies, the broad continuum peaking around 45\,meV has been established 
as a generic, quasi-2D magnetic excitation of the $j$\,=\,1/2 honeycomb iridates 
which persists up to 300\,K \cite{Kim20,Revelli20,Chun21,KimNax22}. A similar continuum 
has been observed in the $j$\,=\,1/2 honeycomb compound $\alpha$-RuCl$_3$ \cite{Suzuki21,Banerjee17}. 
The 10\,meV peak is particularly pronounced at the $K$ point, 
which agrees with first RIXS results collected with improved energy resolution 
$\delta E$\,=\,12\,meV \cite{Kim20}. Observation of the 10\,meV feature with 
$\delta E$\,=\,25\,meV requires excellent suppression of elastic Thomson scattering which we 
achieve by using $2\theta$\,=\,$90^\circ$. For comparison, reference spectra 
(open symbols) measured with $2\theta$\,$\neq$\,$90^\circ$ are peaking at zero energy loss.

In general, RIXS spectra as a function of energy loss are appropriate to study 
dispersive modes. However, for both inelastic features the peak energies hardly 
depend on $\mathbf{q}$, see Fig.~\ref{fig:spectra}. The key to a microscopic 
understanding of the two predominantly local excitations is the 
$\mathbf{q}$-dependent intensity $I(\mathbf{q},\omega)$. This is most evident 
from the astounding behavior at $X_x$ and $X_z$, see Fig.\ \ref{fig:spectra}c). 
These $\mathbf{q}$ points are fully equivalent for the honeycomb lattice 
but probe different bond directions in the case of nearest-neighbor 
excitations, as argued below. We will show that a small value of 
the polarization factor $|P^z(\mathbf{q})|^2$ depicted in 
Fig.~\ref{fig:setup}d) suppresses $j^z$-conserving excitations and enhances 
$j^z$-flip modes. 
This suppresses the spin-conserving 10\,meV feature and enhances the 45\,meV spin-flip 
mode at $X_z$ but not at $X_x$, i.e., on $z$ bonds but not on $x$ bonds. 
Hence $I(\mathbf{q},\omega)$ demonstrates that the magnetic honeycomb lattice 
of Na$_2$IrO$_3$ hosts BDE that have spin-flip or spin-conserving character.

%%%%%%%%%%%%%%%%%%%%%%%%%%%%%%%%%%%%%%%%%%%%%%%%%%%%%%
% RIXS intensity maps
%%%%%%%%%%%%%%%%%%%%%%%%%%%%%%%%%%%%%%%%%%%%%%%%%%%%%%

\begin{figure}[t]
	\centering
	\includegraphics[width=\columnwidth]{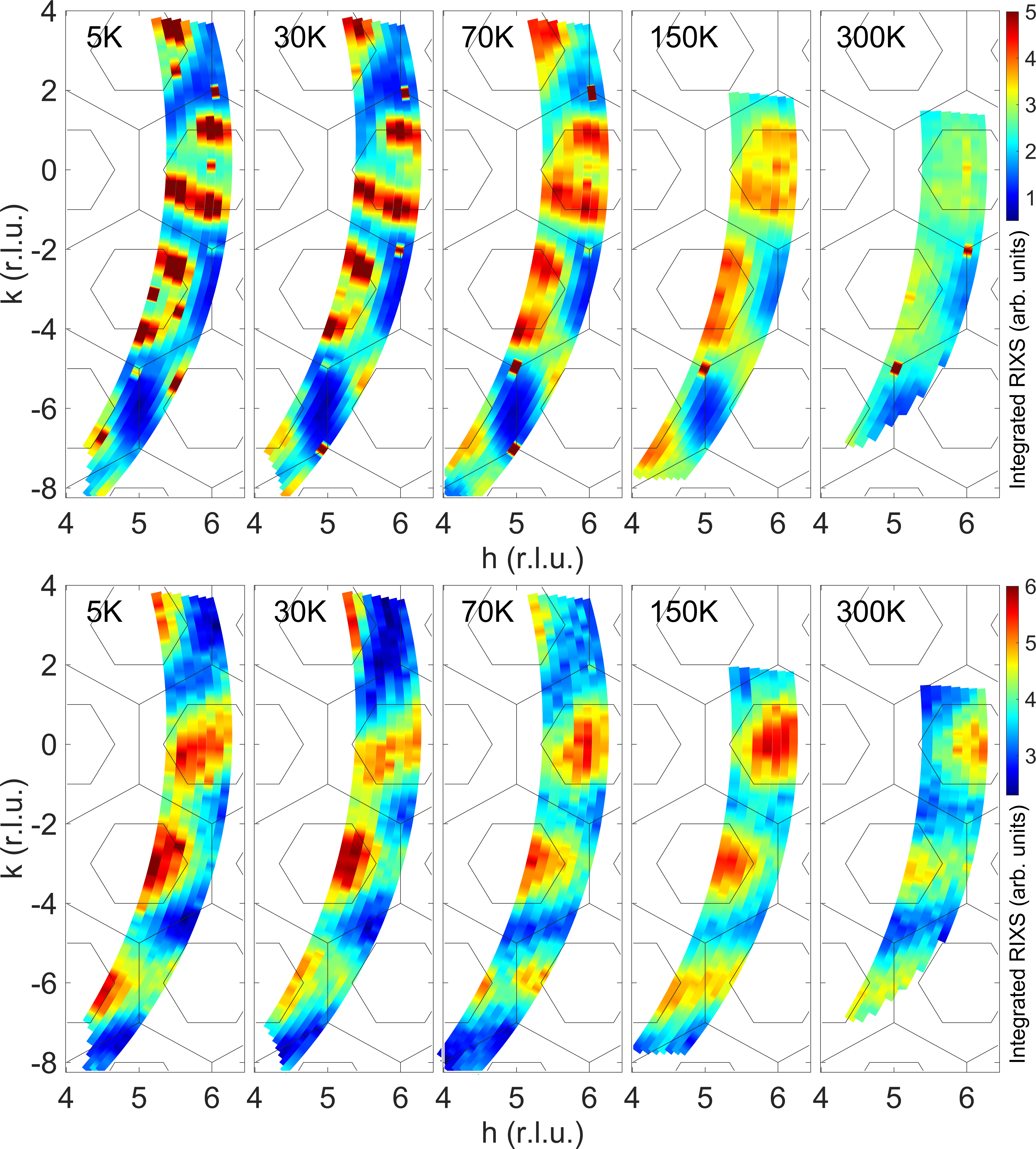}
	\caption{\textbf{RIXS intensity maps at different temperatures,} integrated from 
	-15 to 15\,meV (top) and from 45 to 125\,meV (bottom), focusing on the behavior 
	of the two peaks shown in the spectra, cf.\ Fig.~\ref{fig:spectra}. 
	At low energy, the narrow spots for integer $h$ and $k$ such as at (6\,\,2) 
	correspond to tails of structural Bragg peaks. 
	} 
	\label{fig:map_T}
\end{figure}

{\it RIXS intensity maps:}
Figure \ref{fig:map_T} depicts 2D $\mathbf{q}$-space 
maps of $I(\mathbf{q})$ integrated below 15\,meV and above 45\,meV, respectively. 
The integration ranges have been chosen to disentangle the different 
behavior of the two peaks observed in the spectra. 
For the continuum peaking around 45\,meV, the quasi-2D character is supported by 
the gradual evolution with temperature $T$, see bottom panels in 
Fig.~\ref{fig:map_T}. This insensitivity to $T$ and in particular to $T_N$ 
agrees with previous RIXS results \cite{Revelli20} for selected values of $\mathbf{q}$. 
The overall behavior with a broad peak centered at $\Gamma$ can be described 
by nearest-neighbor correlations (see below).

\begin{figure}[t]
	\centering
	\includegraphics[width=\columnwidth]{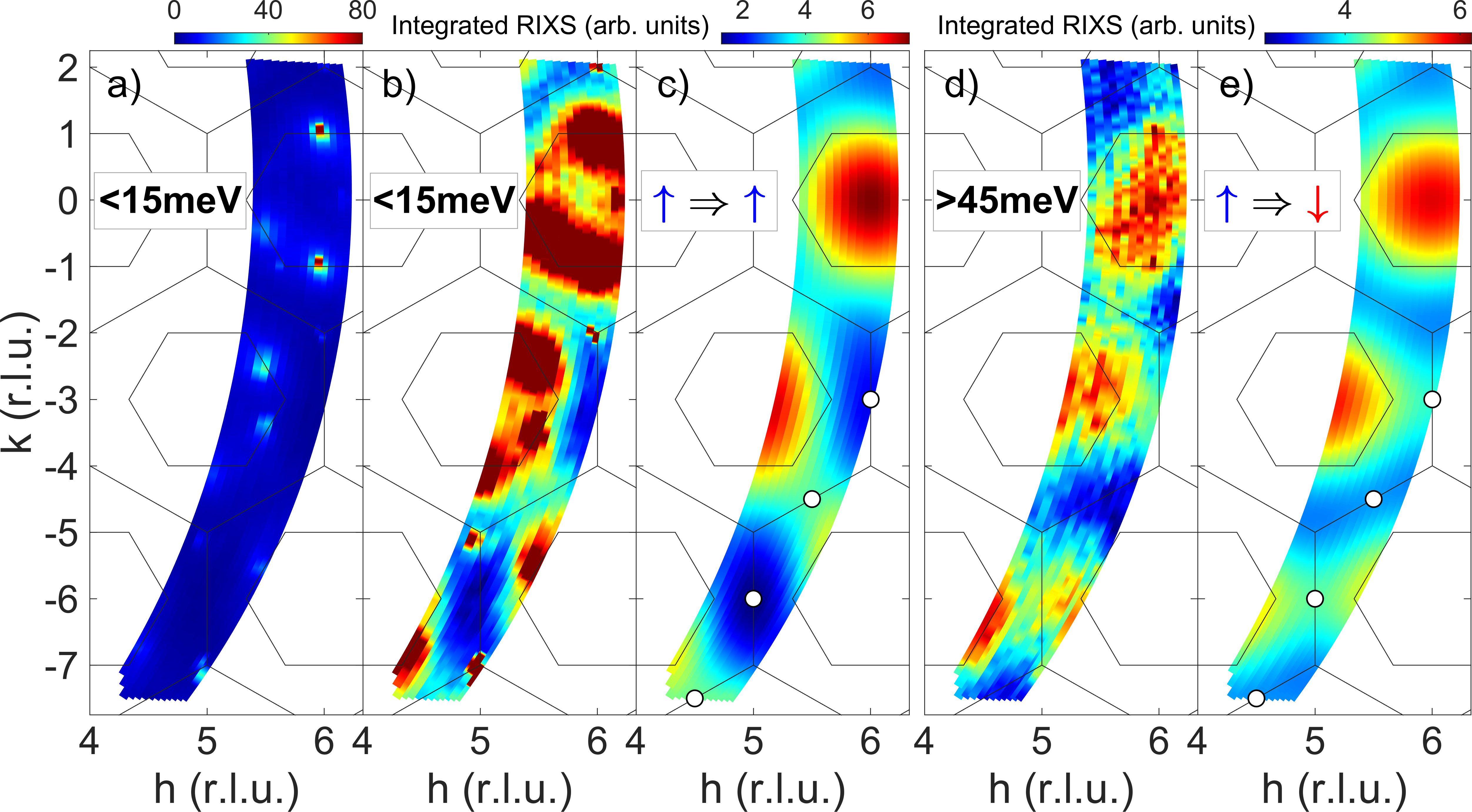}
	\caption{\textbf{Maps of the RIXS intensity} at 5\,K integrated from -15 to 
	15\,meV ((a), (b)) and from 45 to 125\,meV (d). Compared to Fig.~\ref{fig:map_T}, 
	the data were measured on a finer mesh in $\mathbf{q}$ space.		
	Panels (a) and (b) show the same data on different color scales. 
	The bond-directional character is apparent from polarization-related differences 
	of the RIXS intensity at, e.g., the $X_\gamma$ points (white circles). 
	(c), (e): Predictions of a simple bond model for BDE with spin-conserving and 
	spin-flip character, respectively, see Eqs.\ (\ref{eq:I}) and (\ref{eq:Igamma}).
	Excitations are restricted to either an $x$, $y$, or $z$ bond with corresponding 
	correlations of $j^x$, $j^y$, or $j^z$, respectively. The calculated modulation 
	pattern reflects nearest-neighbor correlations, cf.\ Fig.~\ref{fig:mod}.
	} 
	\label{fig:maps}
\end{figure}

To study the 10\,meV peak, we have to cope with the fact that 
the dominant contribution to the low-energy intensity below $T_N$ stems from 
elastic magnetic Bragg scattering with the 3D ordering wave vector 
$\mathbf{Q}_0$\,=\,(0\,\,1\,\,1/2) \cite{Ye12,Choi12} and from low-energy magnons 
that are expected to emerge from there. In the 2D ($h$\,\,$k$) maps, the magnetic 
Bragg spots are not hit perfectly since the value of $l$ is adapted to achieve 
$2\theta$\,=\,90$^\circ$. Nevertheless we find pronounced maxima at 
$M$\,=\,$\Gamma \pm$\,(0\,\,1) and enhanced intensity at $M^\prime$, 
cf.\ Fig.~\ref{fig:setup}c). The narrow features at $M$ and $M^\prime$ are evident 
in Fig.~\ref{fig:maps}a), which shows the same 5\,K data as panel b) but on 
another color scale. 
Chun \textit{et al.} \cite{Chun15} analyzed the elastic scattering at $M$ and $M^\prime$ 
to derive the existence of dominant Kitaev exchange interactions in Na$_2$IrO$_3$. 
In contrast, we focus on the inelastic response not too close to $M$ and $M^\prime$.

With decreasing temperature, the low-energy maps in the top panels of 
Fig.~\ref{fig:map_T} show the building up of intensity close to $M$, reflecting
the evolution of longer-range 3D correlations. However, central to our study is the 
behavior in $\mathbf{q}$ ranges not too close to $M$ such as around the four 
$X_\gamma$ points marked by white symbols in Fig.~\ref{fig:maps}b). 
There, also the low-energy maps are insensitive to $T$, and the data at 5\,K 
and 30\,K\,$\approx 2T_N$ are very similar. In agreement with the spectra, 
cf.\ Fig.~\ref{fig:spectra}c), the low-energy maps show extended ranges of 
low intensity around $X_z$\,=\,(5\,\,-6) and $X_z$\,=\,(6\,\,-3) but larger intensity 
around $X_x$\,=\,(4.5\,\,-7.5) and  $X_x$\,=\,(5.5\,\,-4.5). 
The opposite behavior is observed for the 45\,meV continuum in the high-energy maps, 
again in agreement with the spectra.
The tilted sample geometry of our experiment has been designed to address this 
particular behavior via the polarization factors, revealing a clear signature of 
bond-directional behavior, as we will show below.

Below $T_N$, magnetic order breaks the three-fold rotational symmetry of the 
honeycomb lattice. However, immediately above $T_N$ short-range zigzag correlations 
of all three domains were observed with equal strength \cite{Chun15}. 
Therefore, the inequivalence of $X_x$ and $X_z$ above $T_N$ and 
the corresponding breaking of rotational symmetry cannot be attributed to 
long-range magnetic order. Furthermore, the RIXS intensity at $X_\gamma$ is 
insensitive to $l$, and the intensity maps measured at 5\,K on two different 
magnetic domains are very similar, see \textit{Supplementary Information} \cite{SI}. 
All of these results firmly establish the quasi-2D character of the studied excitations.

%%%%%%%%%%%%%%%%%%%%%%%%%%%%%%%%%%%%%%%%%%%%%%%%%%%%%%
% Prevailing nearest-neighbor correlations
%%%%%%%%%%%%%%%%%%%%%%%%%%%%%%%%%%%%%%%%%%%%%%%%%%%%%%

{\it Nearest-neighbor correlations:}  
From the RIXS intensity maps we now establish the predominance of nearest-neighbor 
correlations, which will allow us to, e.g., identify the response at $X_\gamma$ 
with the $\gamma$ bond. A magnetically ordered state is characterized 
by long-range spin-spin correlations. Above the ordering temperature $T_N$, 
thermal fluctuations yield a strong decrease of the correlation length. 
In quasi-2D compounds, short-range correlations typically survive at temperatures 
much higher than $T_N$ \cite{Ronnow99}. The Kitaev model describes a very 
different case in which spin-spin correlations are strictly restricted to two 
nearest neighbors on a single bond.

\begin{figure}[t]
	\centering
\includegraphics[width=\linewidth]{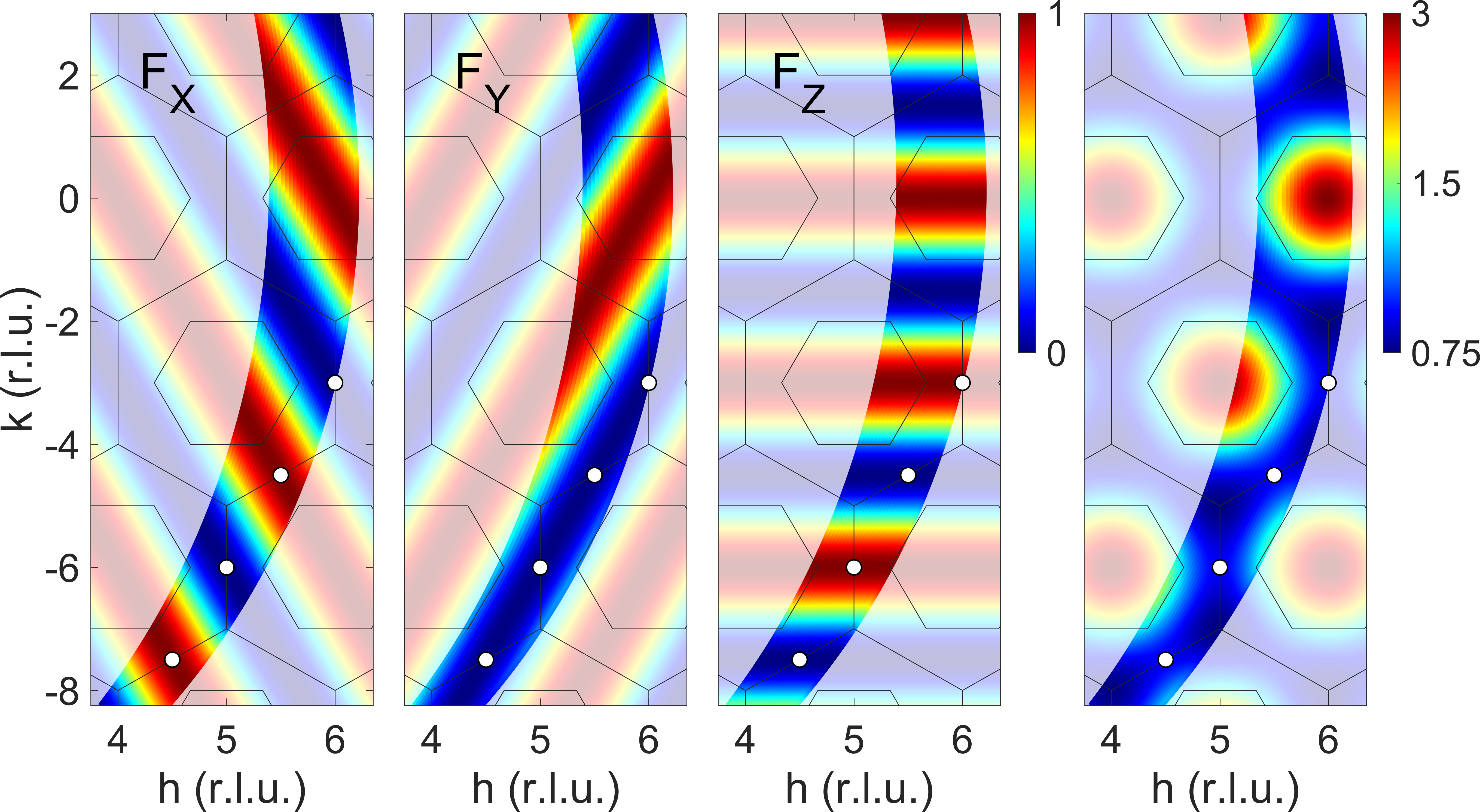}
	\caption{\textbf{Sinusoidal two-site modulation patterns.} The three panels on 
	the left depict the structure factor 
	$F_\gamma$\,=\,$\cos^2(\mathbf{q}\cdot \Delta \mathbf{R}_\gamma/2)$ 
    for individual $\gamma$ bonds with $\gamma$\,=\,$x,y,z$, see Eq.\ \eqref{eq:I}.
    The measured $\mathbf{q}$ range is highlighted, cf.\ Fig.~\ref{fig:map_T}. 
    Right: Sum over $x$, $y$, and $z$ bonds. 
    White circles mark $X_x$ and $X_z$ points, these are relevant for the discussion 
    of the bond-directional character, see Fig.~\ref{fig:spectra}c).
} 
	\label{fig:mod}
\end{figure}

This unusual scenario of nearest-neighbor correlations leaves clear fingerprints 
in the $\mathbf{q}$ dependence of the RIXS intensity \cite{Revelli20}. 
The dynamical structure factor of two sites has sinusoidal shape, i.e., 
the two-site scattering problem is equivalent to an inelastic incarnation of 
Young's double-slit experiment \cite{Revelli19}. 
Summing over the three different bonds one finds 
\begin{equation}
	I_{\rm nn}(\mathbf{q}) = I_0 + \sum_\gamma \,
	I_{\gamma}(\mathbf{q}) \, \cos^2(\mathbf{q}\cdot \Delta\mathbf{R}_{\gamma}/2)  
	\, , 
	\label{eq:I}
\end{equation}
where $\Delta\mathbf{R}_{\gamma}$\,=\,$\mathbf{R}_{2}-\mathbf{R}_{1}$ 
denotes a $\gamma$ bond with two correlated nearest-neighbor Ir sites, 
$I_0$ is a background intensity, and $I_\gamma$ depends on the polarization.  
Such sinusoidal behavior has been observed for the continuum in the Kitaev 
material $\alpha$-RuCl$_3$ in inelastic neutron scattering \cite{Banerjee17}. 
In RIXS on the honeycomb iridates, Eq.\ (\ref{eq:I}) has been found to describe the 
integrated intensity of the continuum along $\Gamma$-$M$-$\Gamma^\prime$ and 
$\Gamma$-$K$-$X$ \cite{Revelli20}, treating $I_0$ and $I_\gamma$ 
as empirical fit parameters. 
In Fig.~\ref{fig:mod}, we visualize the sinusoidal intensity modulation of each bond 
for $I_\gamma$\,=\,1 and $I_0$\,=\,0. 
The sum $I_{\rm nn}(\mathbf{q})$ in the right panel yields broad maxima at $\Gamma$ 
and reduced intensity at $\Gamma^\prime$. 
Comparing $I_{\rm nn}(\mathbf{q})$ with the high-energy maps in the bottom panels 
of Fig.~\ref{fig:map_T} demonstrates that already this simple picture of individual 
bonds describes the behavior of the continuum surprisingly well. 
For integration below 15\,meV, the RIXS intensity in the first Brillouin zone 
is dominated, at low temperature, by longer-range correlations of zigzag type, 
as discussed above. For $\mathbf{q}$ points not too close to $M$ and $M^\prime$, 
a convincing description in terms of the nearest-neighbor model requires considering 
the polarization-dependent matrix elements, as done in Fig.~\ref{fig:maps}c) 
and discussed below.

In such a nearest-neighbor scenario with two-site structure factors 
$F_\gamma$\,=\,$\cos^2(\mathbf{q}\cdot \Delta\mathbf{R}_{\gamma}/2)$, 
the data at $X_\gamma$ selectively probe the response of the $\gamma$ bond. 
For instance $X_z$\,=\,(5\,\,-6) hosts a maximum of $F_z$ with vanishing contributions 
from $x$ and $y$ bonds, see Fig.~\ref{fig:mod}, while $X_{x}$\,=\,(4.5\,\,-7.5) 
shows a maximum of $F_x$ with $F_y$\,=\,$F_z$\,=\,0.

%%%%%%%%%%%%%%%%%%%%%%%%%%%%%%%%%%%%%%%%%%%%%%%%%%%%%%
% Bond-directional excitations
%%%%%%%%%%%%%%%%%%%%%%%%%%%%%%%%%%%%%%%%%%%%%%%%%%%%%%

{\it Bond-directional excitations:}  
With this information on the bond direction, we utilize polarization selection rules 
to address bond-directional behavior, i.e., whether the application of, e.g., 
the local operator $S_i^x$ creates an excitation connected to a specific bond. 
In the Kitaev model, a RIXS excitation on the $\gamma$ bond requires to apply 
$P^\gamma S_i^\gamma$ \cite{Halasz16,Halasz16comment} with the polarization factor 
$\mathbf{P}$\,=\,$i \varepsilon^\prime \times \varepsilon$ and the incident and 
outgoing polarizations $\varepsilon$ and $\varepsilon^\prime$, respectively. 
As shown below, this describes the behavior of the 10\,meV feature but disagrees 
with the different polarization properties of the 45\,meV continuum. Therefore, 
we employ a complementary microscopic approach and start from the RIXS matrix element 
for magnetic excitations of a $j$\,=\,1/2 moment at site $\mathbf{R}_i$ \cite{Ament11}. 
With the outgoing $\pi$ and $\sigma$ polarizations being perpendicular to the 
incident polarization, see Fig.~\ref{fig:setup}, 
this matrix element is proportional to $\mathbf{P} \cdot \mathbf{S_{i}}$, 
where $\mathbf{S_{i}}$ operates within the $j$\,=\,1/2 subspace.
For the two sites of, e.g., a $z$ bond, we consider that $S_i^z$ 
creates a spin-conserving (sc) excitation while $S_i^x$ or $S_i^y$ 
yield a spin flip (sf) of the $z$ component. For $I_\gamma(\mathbf{q})$ 
in Eq.\ (\ref{eq:I}) this yields 
\begin{equation}
	I_{z}^{\rm sc}(\mathbf{q}) \propto |P^z(\mathbf{q})|^2 \,\, \mathrm{and} \,\,\,
	I_{z}^{\rm sf}(\mathbf{q}) \propto |P^x(\mathbf{q})|^2 + |P^y(\mathbf{q})|^2  
	\label{eq:Igamma}
\end{equation}
and equivalent expressions for $x$ and $y$ bonds. We have chosen the experimental 
geometry with the aim to strongly suppress $|P^z(\mathbf{q})|^2$ 
in the lower parts of the maps, see Fig.~\ref{fig:setup}d). 
More precisely, $|P^z|^2 \propto \sin^2(\theta_{\rm in})$, and the angle 
of incidence $\theta_{\rm in}$ is varied from about $5^\circ$ to 65$^\circ$. 
This corresponds to an ($h$\,\,$k$) range where $k$ roughly runs from -8 to 4 
and depends roughly linearly on $\theta_{\rm in}$. 
In contrast, $|P^x(\mathbf{q})|^2$ and $|P^y(\mathbf{q})|^2$ are nearly constant 
and large, see \textit{Supplementary Information} \cite{SI}. 
In the bond-directional scenario described by Eq.\ (\ref{eq:Igamma}), the suppression 
of $|P^z|^2$ switches off spin-conserving excitations on $z$ bonds 
while it reduces the intensity of spin-flip excitations on $x$ or $y$ bonds 
by about a factor 2. 
This explains the stunning difference of the spectra measured at $X_z$ and $X_x$, 
see Fig.~\ref{fig:spectra}c), if we attribute the peaks at 10 and 45\,meV to 
spin-conserving and spin-flip excitations, respectively, and identify $X_\gamma$ 
with the $\gamma$ bond, as appropriate for nearest-neighbor correlations.

Finally, we calculate maps of the RIXS intensity $I_{\rm nn}(\mathbf{q})$ 
expected for BDE, see Eq.\ (\ref{eq:I}). This combines the sinusoidal structure 
factors $F_\gamma$ of the nearest-neighbor model, containing information on 
the bond direction, with the polarization-dependent $I_\gamma(\mathbf{q})$ 
given in Eq.\ (\ref{eq:Igamma}), reflecting the involved spin component. 
Figures \ref{fig:maps}c) and \ref{fig:maps}e) plot $I_{\rm nn}(\mathbf{q})$ for 
spin-conserving and spin-flip excitations, respectively. The excellent agreement 
with the RIXS data in panels b) and d) corroborates the above assignment 
of spin-conserving and spin-flip excitations at low and high energies, respectively. 
The continuum, integrated above 45\,meV, is described very well over the entire range 
of $\mathbf{q}$ and for all studied temperatures. For integration below 15\,meV, 
our nearest-neighbor model represents the RIXS data very well at high $T$ 
such as 70\,K or 150\,K.\@ At 5\,K, it still captures the behavior not too close to 
$M$, while the response in the vicinity of $M$ reflects longer-range correlations. 
Altogether, bond-directional behavior is most evident around $X_\gamma$ points. 
For instance $X_x$ and $X_z$ are equivalent on the honeycomb lattice and within 
the nearest-neighbor model for isotropic $I_\gamma$\,=\,1, see Fig.~\ref{fig:mod}. 
The anisotropic, bond-directional character of the magnetic excitations in 
Na$_2$IrO$_3$ yields very different RIXS intensities around $X_x$ and $X_z$ 
as well as different behavior at low and high energy.

{\it Outlook: }  
In the quest to identify Kitaev materials, the observation of bond-directional excitations 
via our advanced RIXS scheme can play a decisive role in validating, for a given 
compound, the presence of bond-directional interactions. These are appreciated as a 
source of frustration beyond the geometric frustration of non-bipartite lattices
and key to 
the emergence of non-conventional forms of magnetism (such as recently discussed 
for fcc $j$\,=\,$1/2$ Ba$_2$CeIrO$_6$ \cite{Revelli19a}) 
and potentially spin liquid ground states (as hypothesized \cite{Thompson17} 
for $j$\,=\,1/2 pyrochlore Yb$_2$Ti$_2$O$_7$). A first step in this 
direction might be to apply our approach to other 
honeycomb Kitaev materials, but also to $j$\,=\,$1/2$ systems 
which have not attracted primary interest for bond-directional exchanges such as 
Sr$_2$IrO$_4$. 
In further developing our experimental approach a 
natural next step is to validate other forms of bond-directional interactions, 
such as dominating off-diagonal $\Gamma$-interactions \cite{Rau2016,Trebst22},
and their manifestation in bond-directional excitations.

%%%%%%%%%%%%%%%%%%%%%%%%%%%%%%%%%%%%%%%%%%%%%%%%%%%%%%
% Acknowledgments
%%%%%%%%%%%%%%%%%%%%%%%%%%%%%%%%%%%%%%%%%%%%%%%%%%%%%%

\begin{acknowledgments}
We thank A. Revelli for experimental support and useful discussions. 
We gratefully acknowledge the European Synchrotron Radiation Facility (ESRF) and 
the Advanced Photon Source (APS) for providing beam time and technical support. 
Prior to the measurements at ESRF shown here, we had performed a proof-of-principles 
RIXS study at beamline 27-ID at APS for the feasibility of this measurement geometry 
with a large sample tilt angle $\chi$. APS is a U.S. Department of Energy (DOE) 
Office of Science user facility operated for the DOE Office of Science by 
Argonne National Laboratory under Contract No.\ DE-AC02-06CH11357. 
Furthermore, we acknowledge funding from the Deutsche Forschungs\-gemeinschaft 
(DFG, German Research Foundation) via Project numbers 277146847 
(CRC 1238, projects B03, C03), 247310070 (CRC 1143, project A05), 
and 107745057 (TRR 80). 
\end{acknowledgments}

%%%%%%%%%%%%%%%%%%%%%%%%%%%%%%%%%%%%%%%%%%%%%%%%%%%%%%
% Bibliography
%%%%%%%%%%%%%%%%%%%%%%%%%%%%%%%%%%%%%%%%%%%%%%%%%%%%%%

% \bibliography{RIXS}

\begin{thebibliography}{32}
\bibitem [{Hal()}]{Halasz16comment}%
	\BibitemOpen
	\bibfield  {title} {\bibinfo {title} {{Reference \cite{Halasz16}
				distinguishes spin-conserving and non-spin-conserving channels, refering to
				the spin of an electron in the RIXS process. We measure the
				non-spin-conserving channel in direct RIXS, which in terms of the local
				$j$\,=\,1/2 moments contains both $j^{\gamma}$-flip and
				$j^{\gamma}$-conserving excitations}}}\href@noop {} {\ }\BibitemShut
	{NoStop}%
\end{thebibliography}


\end{document}

\title{Supplementary Information
	\\
	RIXS observation of bond-directional nearest-neighbor excitations 
	in the Kitaev material Na$_2$IrO$_3$}

\author{M. Magnaterra}
\author{K. Hopfer}
\affiliation{Institute of Physics II, University of Cologne, 50937 Cologne, Germany}
\author{Ch. J. Sahle}
\affiliation{ESRF, The European Synchrotron, 71 Avenue des Martyrs, CS40220, 38043 Grenoble Cedex 9, France} 
\author{M. Moretti Sala}
\affiliation{Dipartimento di Fisica, Politecnico di Milano, I-20133 Milano, Italy}
\author{G. Monaco}
\affiliation{Dipartimento di Fisica e Astronomia "Galileo Galilei", Universit\`{a} di Padova, Padova, Italy}
\author{J.~Attig}
\author{C. Hickey}
\affiliation{Institute for Theoretical Physics, University of Cologne, 50937 Cologne, Germany}
\author{I.-M. Pietsch}
\author{F. Breitner}
\author{P.~Gegenwart}
\affiliation{Experimental Physics VI, Center for Electronic Correlations and Magnetism, University of Augsburg, 86159 Augsburg, Germany}
\author{M. H. Upton}
\author{Jungho~Kim}
\affiliation{Advanced Photon Source, Argonne National Laboratory, Argonne, Illinois 60439, USA}
\author{S. Trebst}
\affiliation{Institute for Theoretical Physics, University of Cologne, 50937 Cologne, Germany}
\author{P. H. M. van Loosdrecht}
\affiliation{Institute of Physics II, University of Cologne, 50937 Cologne, Germany}
\author{J. van den Brink}
\affiliation{Institute for Theoretical Solid State Physics, IFW Dresden, 01069 Dresden, Germany}
\affiliation{Institute for Theoretical Physics and W\"urzburg-Dresden Cluster of Excellence ct.qmat, Technische Universit\"at Dresden, 01069 Dresden, Germany}
\author{M. Gr\"{u}ninger}
\affiliation{Institute of Physics II, University of Cologne, 50937 Cologne, Germany}

\date{January 19, 2023}

\begin{abstract}
\end{abstract}

\maketitle

\section{Resonance behavior}
\label{sect:resonance}

\begin{figure}[b]
	\centering
	\includegraphics[width=\linewidth]{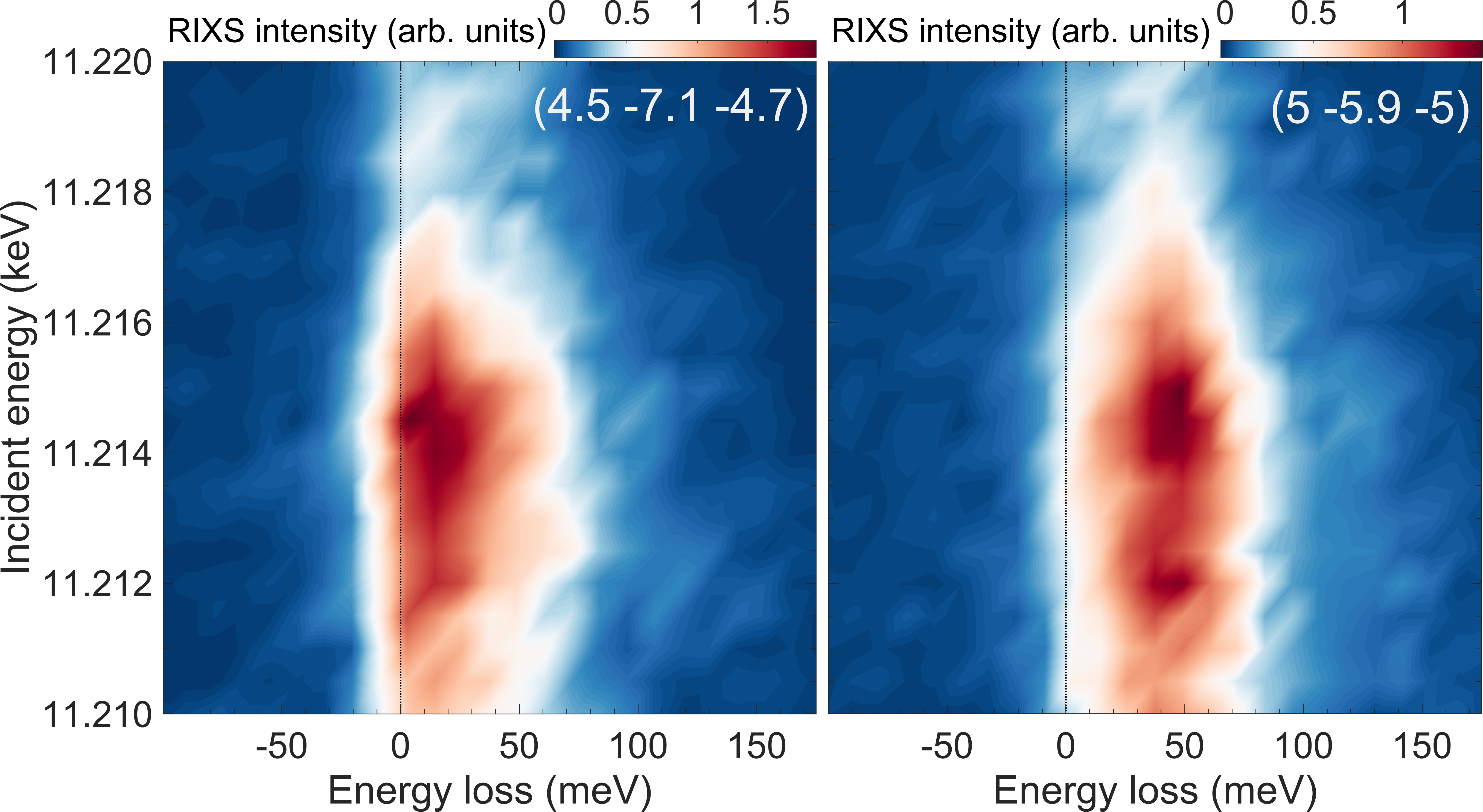}
	\caption{\textbf{Resonance maps} of the magnetic excitations at 30\,K.\@ 
	The 10\,meV peak dominates for transferred momentum \mbox{(4.5\,\,-7.1\,\,-4.7)} 
	(left), while the 45\,meV continuum prevails at \mbox{(5\,\,-5.9\,\,-5)} (right).	
	Both features show $t_{2g}$ resonance at $E_{\rm in}$\,=\,11.2145\,keV 
	(vertical scale). 
	There is no evidence for an $e_g$ resonance at $E_{\rm in}$\,=\,11.218\,keV.\@  
	The elastic line at zero loss (solid black line) is suppressed by using a 
	scattering angle $2\theta$\,=\,$90^\circ$. 
	} 
	\label{fig:res}
\end{figure}

The resonance behavior at the Ir $L_3$ edge provides an unambiguous proof of the 
magnetic nature of the excitations at about 10\,meV and 45\,meV.\@ 
For Ir$^{4+}$ ions, magnetic excitations within the $j$\,=\,1/2 subspace are resonantly 
enhanced in a direct RIXS process that involves the promotion of a $2p$ core electron 
to the $t_{2g}$ orbitals \cite{Ament11RMP}. 
For Na$_2$IrO$_3$, this $t_{2g}$ resonance occurs for an incident energy 
$E_{\rm in}$\,=\,11.2145\,keV \cite{Revelli20}. For comparison, the RIXS intensity 
of excitations to $e_g$ orbitals in the iridates is maximized by choosing a 3 to 4\,eV 
larger value of $E_{\rm in}$ \cite{Lefrancois16}. 
We observe pronounced $t_{2g}$ resonance behavior both around 10\,meV and around 45\,meV, 
see Fig.\ \ref{fig:res}. 
The right panel shows a resonance map measured at (5\,\,-5.9\,\,-5) where the 
45\,meV continuum is pronounced while the 10\,meV peak is suppressed, 
as demonstrated by the RIXS spectra in Fig.\ 2c) of the main text. 
In contrast, the data in the left panel of Fig.~\ref{fig:res} were measured 
at (4.5\,\,-7.1\,\,-4.7) and show a dominant contribution of the 10\,meV feature. 
The common $t_{2g}$ resonance behavior firmly corroborates the magnetic character of 
both excitations.

In contrast, the excitation of phonons corresponds to an \textit{indirect} RIXS process 
that is boosted if $e_g$ orbitals are involved \cite{Moser15}. However, phonons have 
not been detected in $L_3$ edge RIXS data of Mott-insulating $5d^5$ iridates with 
Ir$^{4+}$ ions, see, e.g., Refs.\ 
\cite{KimNatComm14,Lefrancois16,MorettiSr3,Lu17,Rossi17,Revelli19a,Revelli20,Kim20,Chun21}.
The suppression of the phonon contribution can be explained by the well-screened 
and short-lived intermediate state Ir $2p^5$ $t_{2g}^6$ at the $L_3$ edge. 
This has to be distinguished from the case of RIXS at the O $K$ edge with 
$E_{\rm in}$\,$\approx$\,$0.53$\,keV, where the observation of phonon features 
in $\alpha$-Li$_2$IrO$_3$ \cite{Vale19} can be attributed to the very different, 
long-lived intermediate state.

\section{2D character of excitations and insensitivity to magnetic domains}

\subsection{$l$ dependence}

\begin{figure}[t]
 	\centering
 	\includegraphics[width=0.45\linewidth]{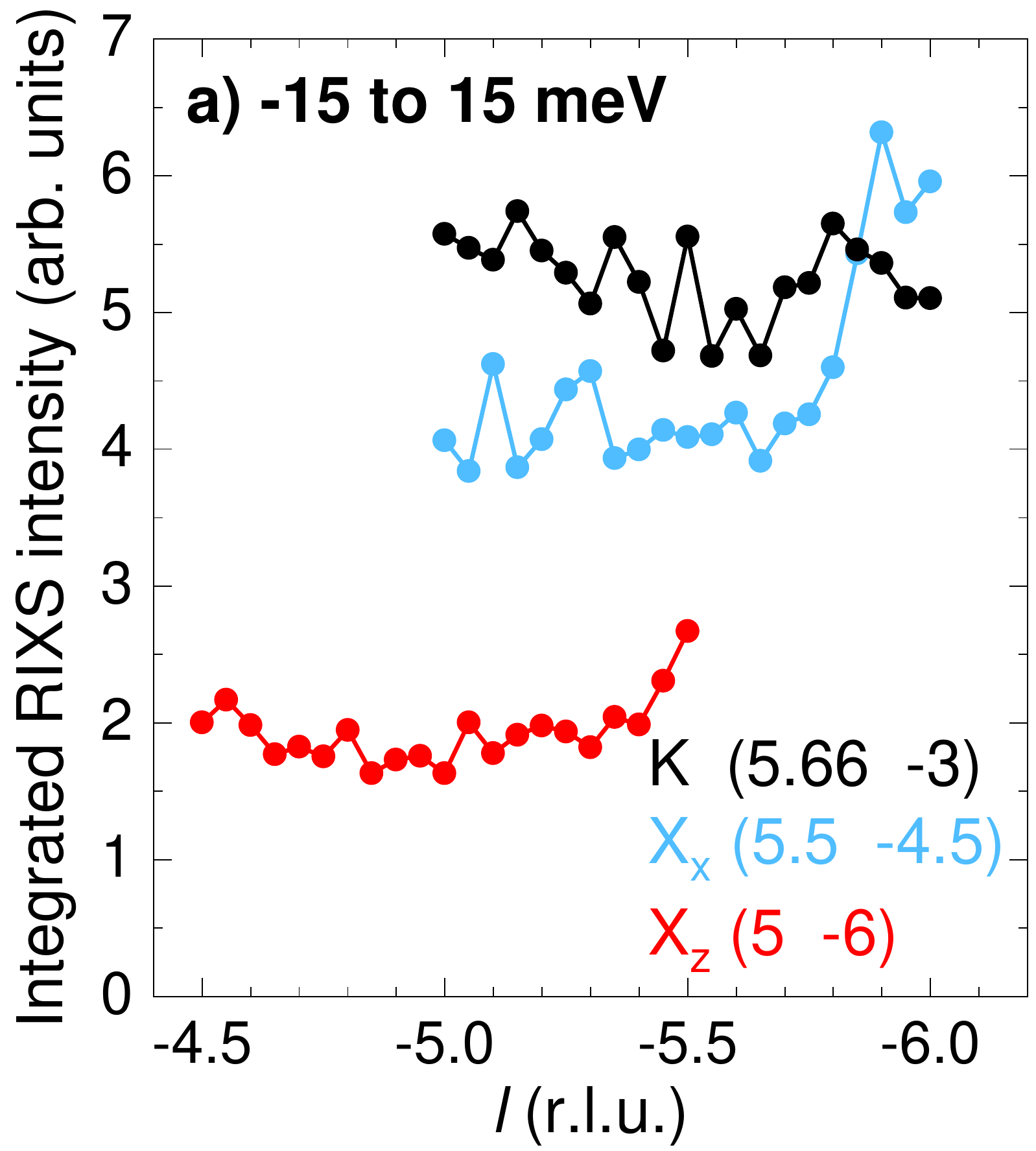}
 	\hspace*{7mm}
 	\includegraphics[width=0.45\linewidth]{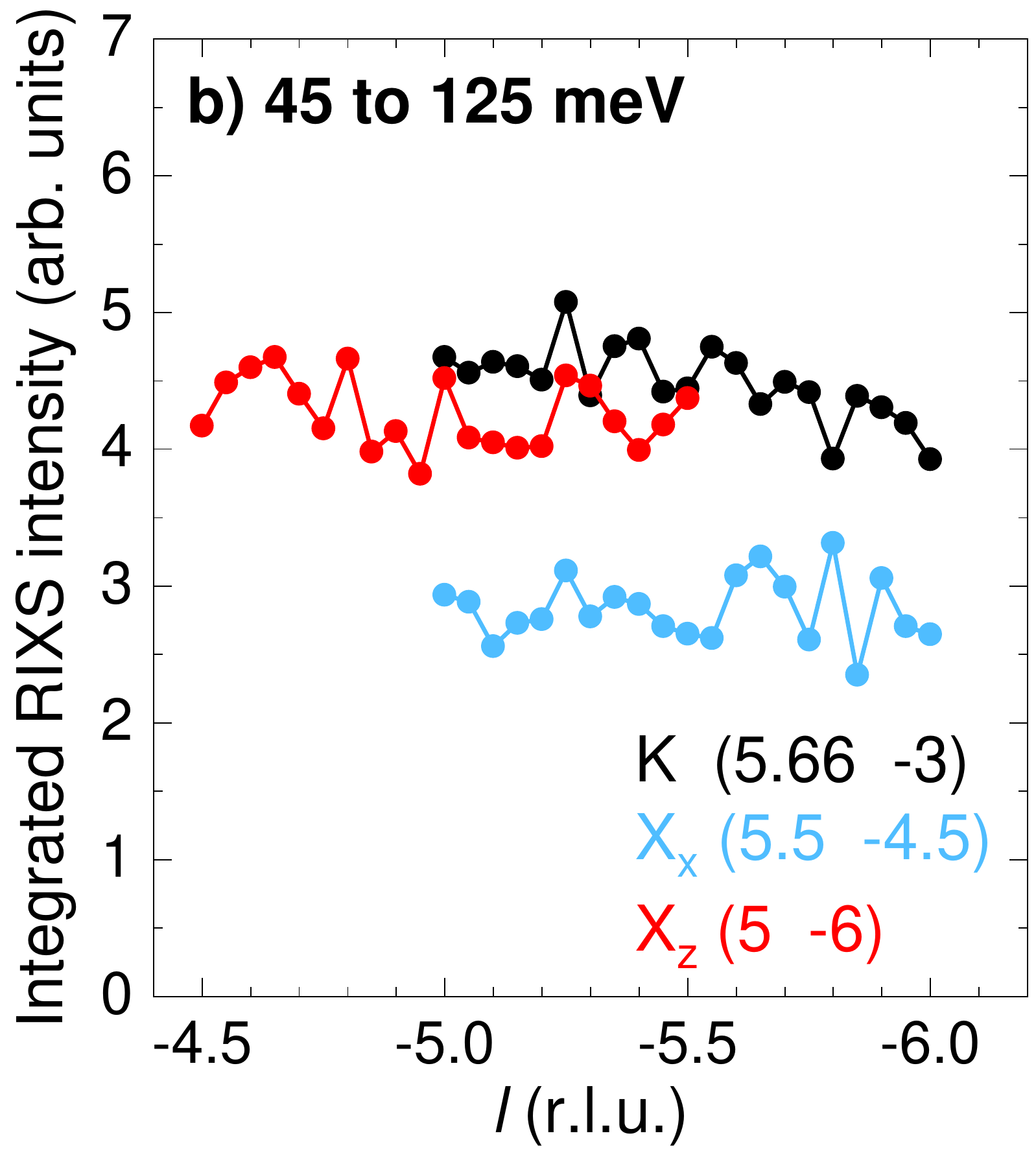}
 	\includegraphics[width=0.45\linewidth]{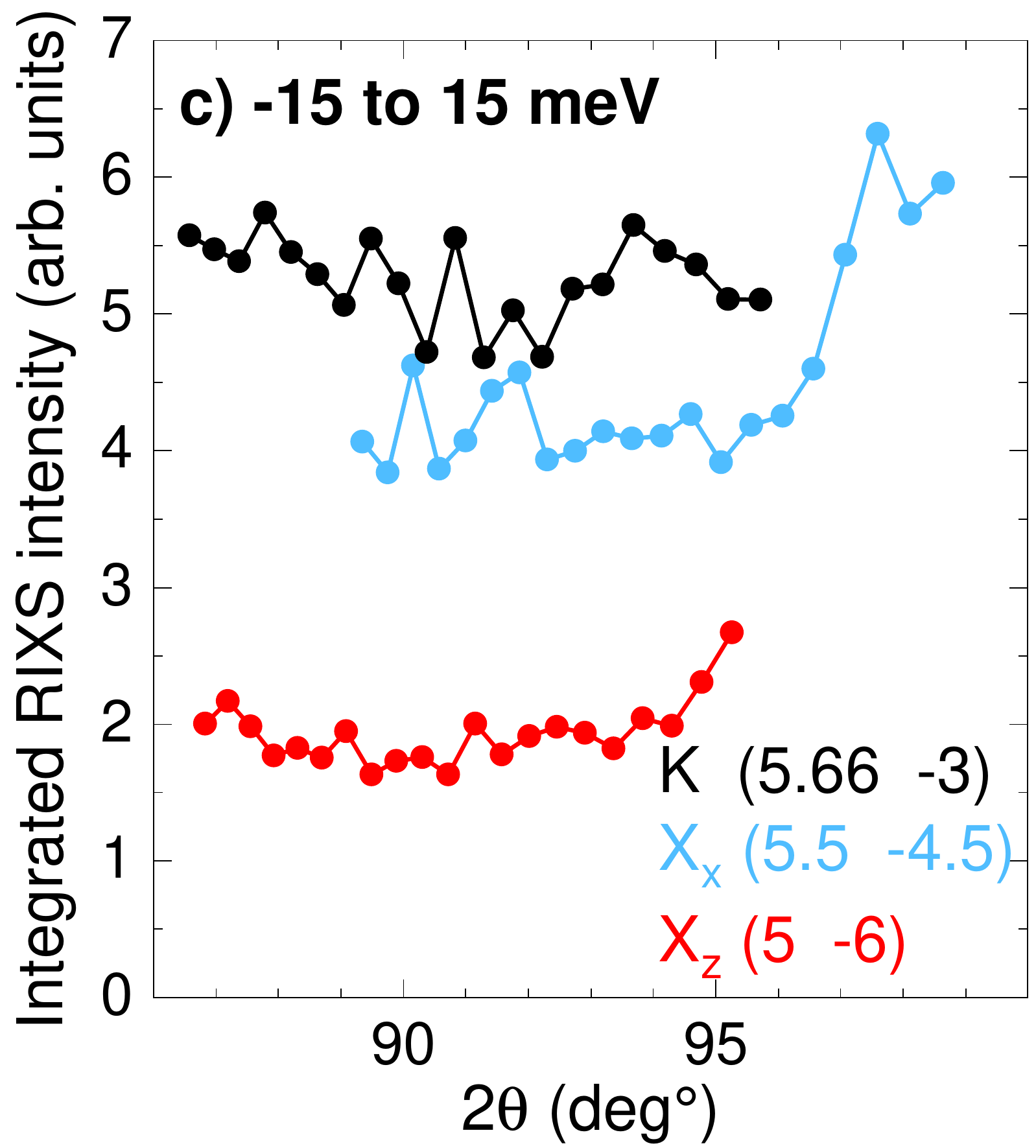}
	\hspace*{7mm}
 	\includegraphics[width=0.45\linewidth]{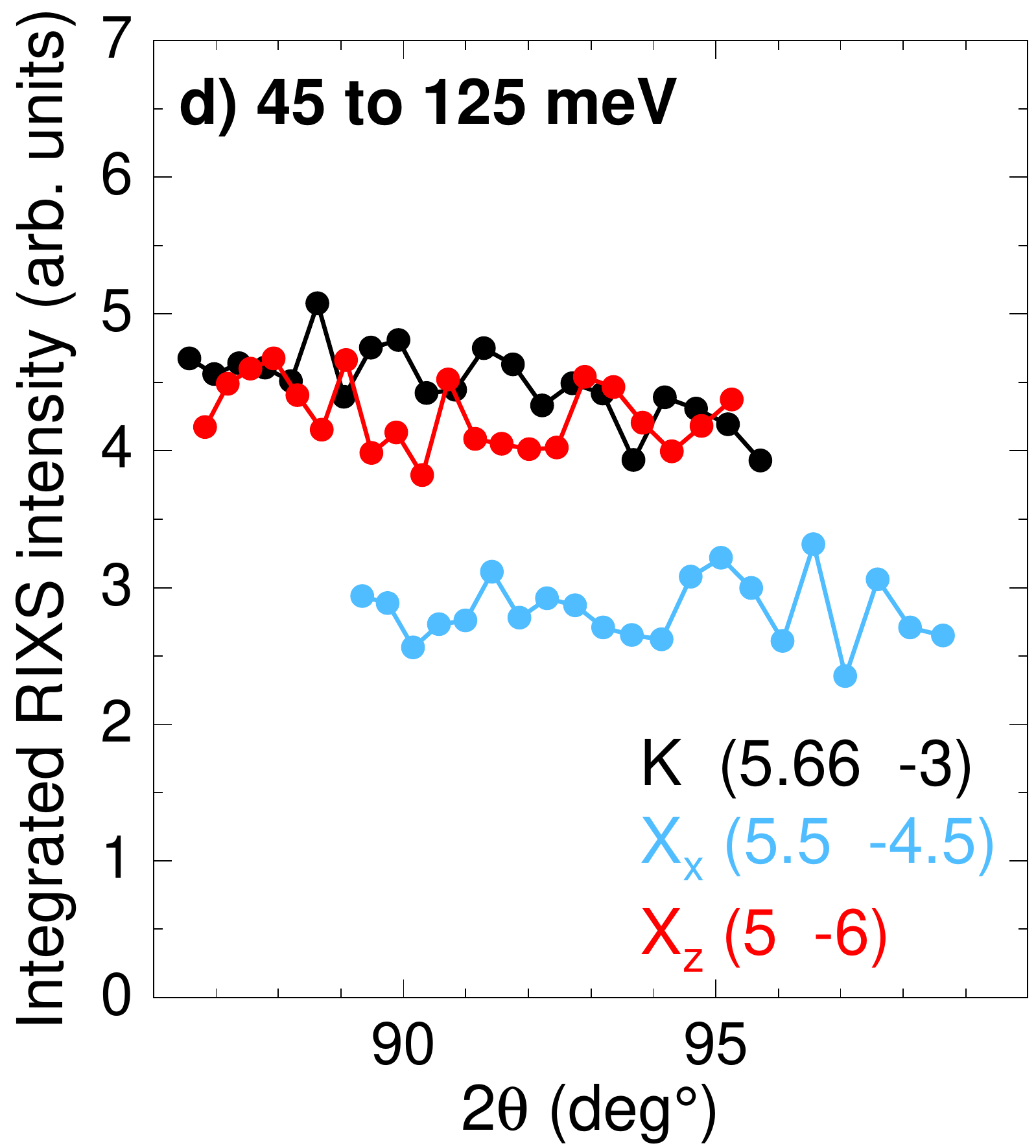}
 	\caption{\textbf{2D character of magnetic excitations.}  
 	Panels a) and b) depict the RIXS intensity at (5.66\,\,-3\,\,$l$), (5.5\,\,-4.5\,\,$l$), 
 	and (5\,\,-6\,\,$l$) for integration below 15\,meV and above 45\,meV, respectively. 
 	The data have been collected at 5\,K.\@ 
 	Panels c) and d) show the same data plotted as a function of the scattering angle $2\theta$. 
 	At low energy, the contribution of the elastic line is suppressed for $2\theta$\,=\,$90^\circ$.
 	} 
 	\label{fig:L}
\end{figure}

Below the N\'eel temperature $T_N$\,=\,15\,K, Na$_2$IrO$_3$ hosts long-range magnetic 
order with the 3D propagation vector $\mathbf{Q}_0$\,=\,(0\,\,1\,\,1/2) \cite{Ye12,Choi12}. 
In the 3D ordered state, low-energy magnons are expected to emerge from $\mathbf{Q}_0$ 
and to show a dispersion as a function of the transferred momentum $\mathbf{q}$. 
However, we focus on the magnetic excitations at about 10\,meV and 45\,meV that 
cannot be described as magnons of the long-range ordered phase. 
These features persist to temperatures far above $T_N$, see Fig.\ 3 of the main text, 
which provides strong evidence for a predominantly 2D character. Our study highlights 
the RIXS intensity of these 2D excitations in ($h$\,\,$k$) space not too close 
to $M$\,=\,($0$\,\,$\pm 1$) and $M^\prime$\,=\,($\pm 1/2$\,\,$\pm 1/2$), 
cf.\ Fig.\ 1c) in the main text. 
In the ($h$\,\,$k$) range relevant to us, the RIXS intensity is insensitive to $l$ 
even at 5\,K, which is demonstrated in Fig.\ \ref{fig:L}, 
using $X_z$\,=\,(5\,\,-6), $X_x$\,=\,(5.5\,\,-4.5), and $K$\,=\,(5.66\,\,-3) as examples. 
Panels a) and b) show data as a function of $l$ for integration of the RIXS intensity 
below 15\,meV and above 45\,meV.\@ In the latter case, the RIXS intensity is roughly 
constant as a function of $l$. In particular, the intensity of the 45\,meV feature 
is significantly lower at $X_x$ than at $X_z$ for all studied values of $l$. 
The 15\,meV shows the opposite behavior, the RIXS intensity is higher at $X_x$ than 
at $X_z$, again for all $l$. For integration from -15 to 15\,meV, 
the contribution of the elastic line has to be considered. In the data shown in 
Figs.\ 2, 3, and 4 of the main text, the elastic line has been suppressed by choosing 
a scattering angle $2\theta$\,=\,$90^\circ$. This, however, fixes $l$ for a given 
($h$\,\,$k$) point. Measuring the $l$ dependence requires to change $2\theta$ away 
from $90^\circ$. Therefore, we replot the same data in panels c) and d) as a function 
of $2\theta$. Close to $90^\circ$, the RIXS intensity is roughly constant also 
for integration below 15\,meV, which strongly corroborates the 2D character. 
The increase of the low-energy intensity for $2\theta$ further away from $90^\circ$ 
arises from the increasing elastic contribution of Thomson scattering.

\subsection{3D magnetic domains}

The 2D character is further supported by the insensitivity of our results to 3D magnetic domains. 
The ideal 2D honeycomb lattice exhibits threefold rotational symmetry, such that 
the $M$ points ($0$\,\,$\pm 1$) and the $M^\prime$ points ($\pm 1/2$\,\,$\pm 1/2$) 
all are equivalent.  
However, the 3D crystal structure of Na$_2$IrO$_3$ shows a small monoclinic distortion, 
hence ($0$\,\,$\pm 1$\,\,$1/2$) and ($\pm 1/2$\,\,$\pm 1/2$\,\,$1/2$) are not 
equivalent.
With the 3D propagation vector $\mathbf{Q}_0$\,=\,(0\,\,1\,\,1/2) of long-range 
magnetic order \cite{Ye12,Choi12}, 
the orientation of the zigzag ordering pattern is tied to the crystal structure.

\begin{figure}[t]
\centering
\includegraphics[width=\linewidth]{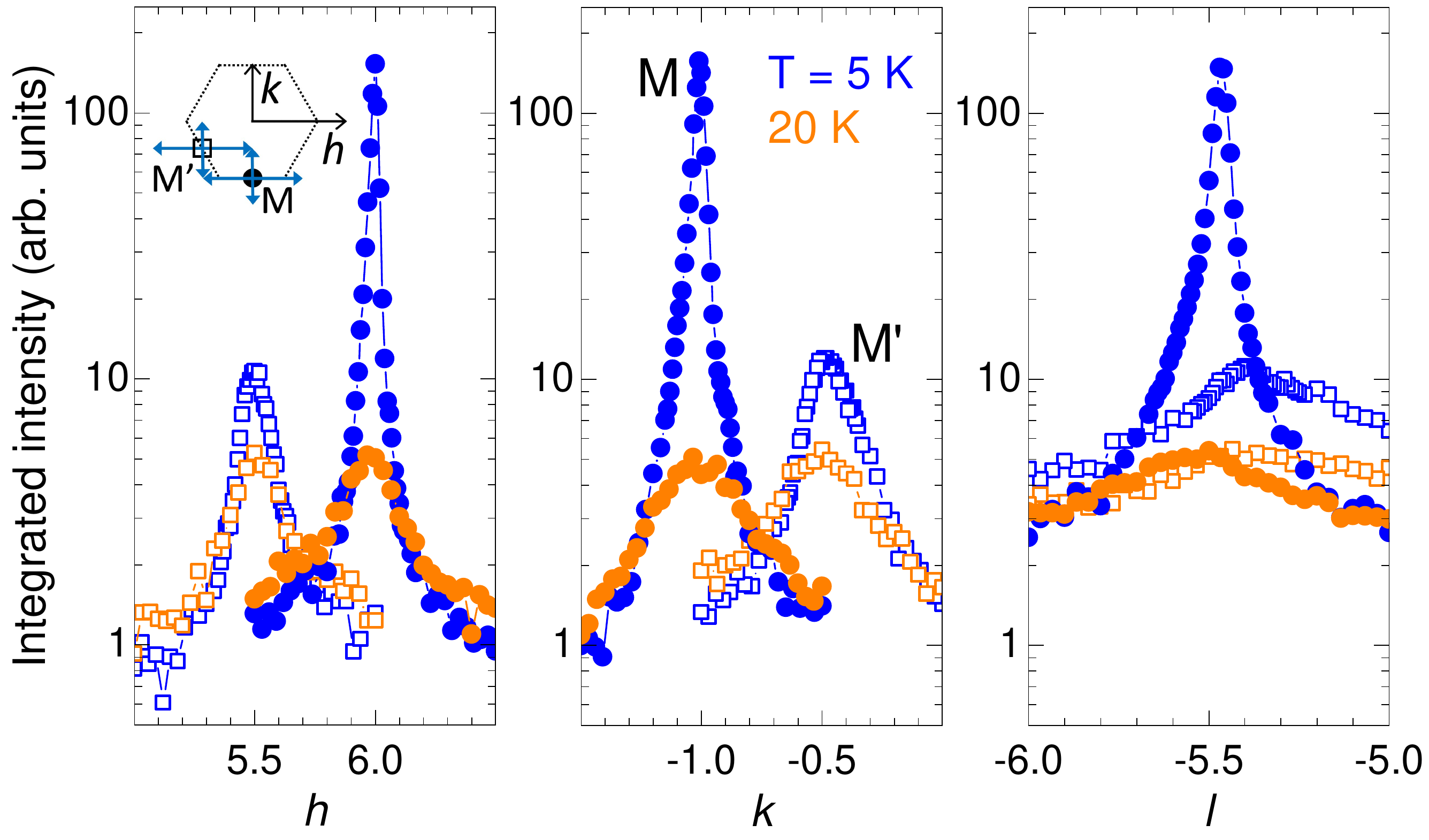}
	\caption{\textbf{Quasielastic scans across a magnetic Bragg peak.} 
	The scattered intensity was integrated from -15\,meV to +15\,meV.\@ 
	Full and open symbols show data across (6\,\,-1\,\,-5.5) and (5.5\,\,-0.5\,\,-5.5), 
	respectively. At 20\,K, above $T_N$, both $\mathbf{q}$ points show the same intensity. 
	At 5\,K, (6\,\,-1\,\,-5.5) hosts a magnetic Bragg peak. 
	} 
	\label{fig:Bragg}
\end{figure}

Laue diffraction of our samples indicates structural $120^\circ$ twinning in 
the honeycomb plane. The structural twins dictate the formation of corresponding 
magnetic domains. The cross section of the incident x-ray beam of roughly 
$(15\times 15)$\,$\mu$m$^2$ allows us to select the measurement spot such that 
a given twin domain dominates the response, as demonstrated by scans across 
$M_{\rm 3D}$\,=\,(6\,\,-1\,\,-5.5) and $M^\prime_{\rm 3D}$\,=\,(5.5\,\,-0.5\,\,-5.5) 
for zero energy loss, see Fig.\ \ref{fig:Bragg}. 
At 20\,K, i.e., above $T_N$, we observe clear peaks of the elastically scattered intensity 
along $h$ and $k$ but a broad intensity distribution along $l$. Furthermore, the intensity 
is very similar at $M_{\rm 3D}$ and $M^\prime_{\rm 3D}$. This suggests the coexistence of 
short-range 2D zigzag fragments running along three equivalent directions, in agreement 
with the results of Chun \textit{et al.} \cite{Chun15}. 
At 5\,K, we find a magnetic Bragg peak at $M_{\rm 3D}$ that is characterized by 
a pronounced peak as a function of $l$. 
In contrast, the intensity at $M^\prime_{\rm 3D}$ is much lower and the peak still 
is very broad as a function of $l$. 
This demonstrates that the measurement predominantly probes a single magnetic domain.

\begin{figure}[t]
	\centering
	\includegraphics[width=\linewidth]{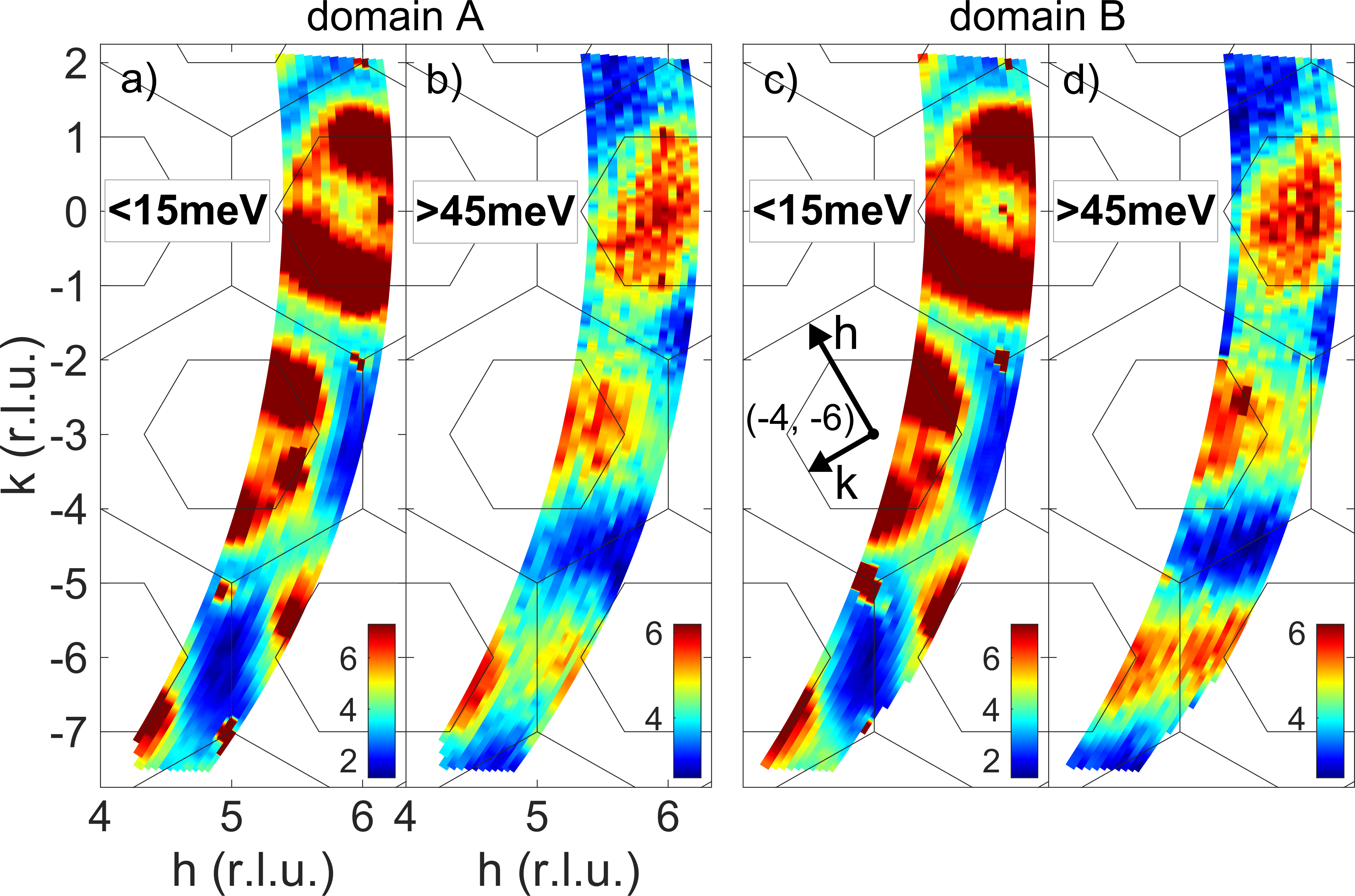}
	\caption{\textbf{Maps of the RIXS intensity of two twin domains} measured at 5\,K.\@ 
	Domain B is rotated by $120^\circ$ with respect to domain A, as indicated by the 
	coordinate frame in panel c). The intensity has been integrated from -15 to 15\,meV and 
	from 45 to 125\,meV.\@ The data agree on the bond-directional character of the 
	magnetic excitations.
	} 
	\label{fig:domain}
\end{figure}

In Fig.\ \ref{fig:domain} we compare ($h$\,\,$k$) maps of the RIXS intensity 
measured at 5\,K on two different spots of the sample that correspond to two different 
twin domains rotated by $120^\circ$. For each spot, we study the intensity integrated 
from -15 to 15\,meV and from 45 to 125\,meV.\@ 
The intensity maps for the two different domains are very similar. 
In particular, the distinct behavior at different $X_\gamma$ points is observed 
equivalently. This result corroborates that the different intensities at $X_x$ and $X_z$ 
cannot be attributed to small differences arising from the 3D crystal structure  
but reflect bond-directional excitations that are detected based on the polarization dependence. 
Note that the unambiguous determination of the character of a specific domain requires 
to address specific ($h$\,\,$k$\,\,$l$) points, the domain character cannot be read 
from ($h$\,\,$k$) maps measured with $2\theta$\,=\,$90^\circ$.

\section{Polarization factors}

\begin{figure}[t]
	\centering
	\includegraphics[width=\linewidth]{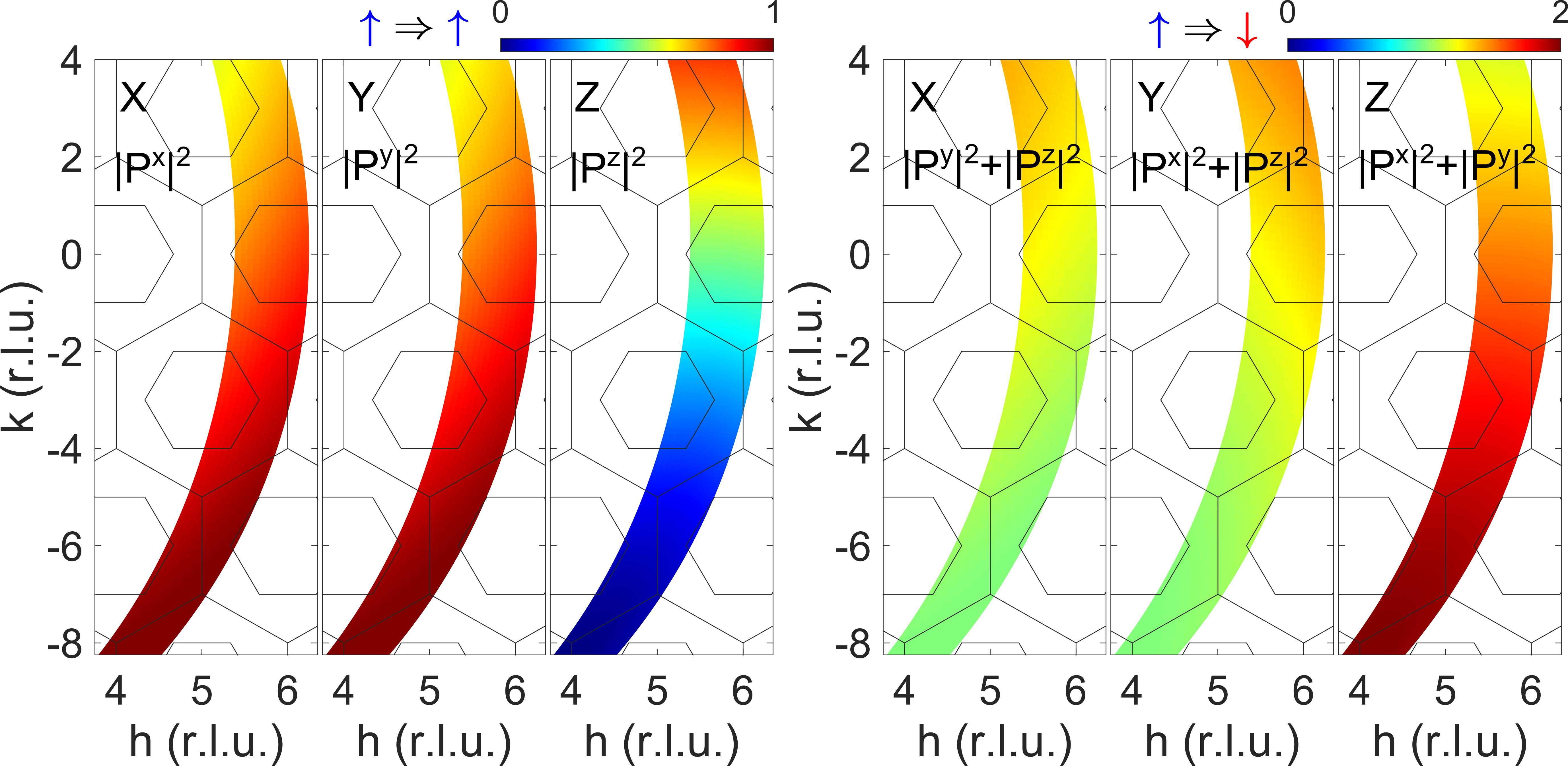}
	\caption{\textbf{Polarization factors} of the three bonds. 
		Left: $|P^\gamma(\mathbf{q})|^2$, applicable to spin-conserving excitations 
		on a $\gamma$ bond, as observed at 10\,meV, cf.\ Eq.\ (\ref{eq:Isc}). 
		The central feature is the suppression of $|P^z(\mathbf{q})|^2$ in the 
		lower part of the map, while $|P^x|^2$ and $|P^y|^2$ do not vary strongly 
		with $\mathbf{q}$.
		Right: The panel for the $z$ bond depicts $|P^x(\mathbf{q})|^2 + |P^y(\mathbf{q})|^2$, 
		which is applicable to spin-flip excitations on the $z$ bond, cf.\ Eq.\ (\ref{eq:Isf}). 
		The other panels show equivalent expressions for $x$ and $y$ bonds. 
		This describes the behavior of the 45\,meV feature.
	} 
	\label{fig:pol_factors}
\end{figure}

At the Ir $L$ edge, the RIXS intensity $I(\mathbf{q},\omega)$ for excitations from the 
ground state $|0\rangle$ to a final state $|f\rangle$ is given by a coherent sum of 
scattering processes localized on individual sites. The sum is running over all sites 
$\mathbf{R}_i$ at which a given final state $|f\rangle$ can be reached, i.e., 
over which $|f\rangle$ is delocalized. 
In dipole approximation one finds \cite{Ament11RMP}
\begin{equation}
	I(\mathbf{q},\omega) = \sum_f \left| \langle f |\, \sum_{i} e^{i\mathbf{q}\cdot\mathbf{R_i}}\, 
	\left[ D^\dagger D \right]_{\mathbf{R}_i} \,| 0\rangle \right|^2 \delta(\hbar \omega - E_f)
	\label{eq:RIXS_intensity_general}
\end{equation}
where $E_f$ denotes the excitation energy and $D$ is the local dipole operator. 
In general, magnetic excitations can be generated via different polarization 
channels \cite{Haverkort10}. 
We use $2\theta$\,=\,90$^\circ$ for the scattering angle, hence the outgoing vertical 
and horizontal polarizations $\varepsilon_{\sigma}^\prime$ and $\varepsilon_{\pi}^\prime$ 
are perpendicular to the linear incident polarization $\varepsilon$. In this case, 
the single-site RIXS matrix elements for magnetic excitations within the $j$\,=\,1/2 manifold 
at site $\mathbf{R}_i$ are given by \cite{Ament11}
\begin{equation}
	[D^\dagger D]_{\mathbf{R}_i} = \frac{2}{3} \mathbf{P} \cdot \mathbf{S}_{i} 
	\label{eq:pj}
\end{equation}
with the polarization factor 
$\mathbf{P}$\,=\,$i \varepsilon^\prime \times \varepsilon$, 
and $\mathbf{S}_{i}$ operates within the $j$\,=\,1/2 subspace. 
In the employed experimental geometry with $2\theta$\,=\,90$^\circ$, see Fig.\ 1 of 
the main text, the polarization factor mainly depends on the angle of incidence 
$\theta_{\rm in}$, which translates to a $\mathbf{q}$ dependence of the polarization 
factor.

In terms of the diagonal components of the dynamical spin structure factor $S(\mathbf{q},\omega)$, 
\begin{align}
	S^{\gamma \gamma}(\mathbf{q},\omega) & =  
	\sum_f \left| \langle f |\, \sum_{i} e^{i\mathbf{q}\cdot\mathbf{R}_i}\, S_{i}^\gamma 
	\,| 0\rangle \right|^2 \delta(\hbar \omega - E_f)
	\, ,
	\label{eq:Sqw}
\end{align}
the RIXS intensity can be written as 
\begin{align}
	I(\mathbf{q},\omega)  & \propto \sum_{\gamma \in \{x,y,z\}}  
	|P^\gamma(\mathbf{q})|^2  S^{\gamma \gamma}(\mathbf{q},\omega) 
	\, ,
	\label{eq:ISqw}
\end{align}
where, as usual, we neglect the off-diagonal components of $S(\mathbf{q},\omega)$. 
In this form, the RIXS intensity thus measures the sum of the dynamical spin structure 
factor components weighted by their respective momentum-dependent polarization factors. 
We have chosen the experimental geometry such that $|P^z(\mathbf{q})|^2$ strongly 
differs from the other components, see Fig.\ \ref{fig:pol_factors}. 
By varying $\mathbf{q}$, one can thus vary the weighting of the different components 
of $S(\mathbf{q},\omega)$, e.g.\ suppressing the contribution of $S^{zz} (\mathbf{q},\omega)$.

Considering only nearest-neighbor correlations, the $\mathbf{q}$ 
dependence of the dynamical structure factor is captured using only 
$\cos^2(\mathbf{q} \cdot \mathbf{\Delta R}_\gamma/2)$ terms. 
For the 10\,meV peak (spin-conserving excitations), the total intensity is well described by 
\begin{equation}
	I_{\rm nn}^{\rm sc}(\mathbf{q})  = 
	\sum_\gamma |P^\gamma(\mathbf{q})|^2 \, \cos^2(\mathbf{q} \cdot \mathbf{\Delta R}_\gamma/2)
	\, , 
	\label{eq:Isc}
\end{equation}
where the $\mathbf{q}$ dependence of the structure factor is given by
\begin{equation}
	S^{\gamma\gamma} (\mathbf{q},10\,{\rm meV})  \propto 
	\cos^2(\mathbf{q} \cdot \mathbf{\Delta R}_\gamma/2)
	\, ,
\end{equation}
meaning that $j^\gamma$ correlations dominate on $\gamma$ bonds. 
For the 45\,meV peak (spin-flip excitations), our data agree with 
\begin{eqnarray}
	\label{eq:Isf}
	I_{\rm nn}^{\rm sf}(\mathbf{q}) & = & 
	\sum_\gamma \left[ \sum_{\gamma^\prime \neq \gamma}  |P^{\gamma^\prime}(\mathbf{q})|^2 \right] 
	\, \cos^2(\mathbf{q} \cdot \mathbf{\Delta R}_\gamma/2)
\\
\nonumber
	& = & \sum_\gamma |P^\gamma(\mathbf{q})|^2 
	\left[ \sum_{\gamma^\prime \neq \gamma}  \cos^2(\mathbf{q} \cdot \mathbf{\Delta R}_{\gamma^\prime}/2) \right] 
	\, ,
\end{eqnarray}
which yields  
\begin{equation}
	S^{\gamma\gamma} (\mathbf{q},45\,{\rm meV})  \propto  
	\sum_{\gamma^\prime \neq \gamma} \cos^2(\mathbf{q} \cdot \mathbf{\Delta R}_{\gamma^\prime}/2)
	\, .
\end{equation}
This means that, e.g., application of $S_i^x$ creates a spin flip 
on either a $y$ bond or a $z$ bond. 
From Eqns.~(\ref{eq:Isc}) and (\ref{eq:Isf}) one can read off $I_\gamma^{\rm sc}$ 
and $I_\gamma^{\rm sf}$ used in Eqns.\ (1) and (2) of the main text.

%%%%%%%%%%%%%%%%%%%%%%%%%%%%%%%%%%%%%%%%%%%%%%%%%%%%%%
% Bibliography
%%%%%%%%%%%%%%%%%%%%%%%%%%%%%%%%%%%%%%%%%%%%%%%%%%%%%%

%\bibliography{RIXS}
